\documentclass[preprint,authoryear,10pt,letterpaper]{elsarticle}
\usepackage[top=0.75in, bottom=0.75in, left=0.75in, right=0.75in]{geometry}

\usepackage{amsmath,amsthm}
\usepackage{amssymb}
\usepackage{amsfonts}
\usepackage{graphicx}
\usepackage{subfigure}
\usepackage{algorithm}
\usepackage{algorithmic}
\usepackage{tabularx,booktabs}

\usepackage{tablefootnote}
\usepackage{threeparttable}

\usepackage{xcolor}
\definecolor{darkblue}{rgb}{0.0,0.5,0.5}
\definecolor{blue}{rgb}{0.0,0.5,0.68}
\usepackage[colorlinks]{hyperref}
\hypersetup{colorlinks,breaklinks,linkcolor=darkblue,urlcolor=darkblue,anchorcolor=darkblue,citecolor=darkblue}
\usepackage{booktabs,caption}
\usepackage{multirow}
\usepackage{lscape}
\usepackage{epstopdf}
\usepackage{lineno}
\usepackage{subfigure}
\usepackage{multirow}
\usepackage{makecell}

\usepackage{microtype}
\usepackage{lineno}
\usepackage{tikz}
\newcommand\tikzmark[2]{%
\tikz[remember picture,baseline] \node[inner sep=2pt,outer sep=0] (#1){#2};%
}

\newcommand\link[2]{%
\begin{tikzpicture}[remember picture, overlay, >=stealth, shift={(0,0)}]
  \draw[->] (#1) to (#2);
\end{tikzpicture}%
}

\newcommand\dashlink[2]{%
\begin{tikzpicture}[remember picture, overlay, >=stealth, shift={(0,0)}]
  \draw[->] (#1) to (#2) [dashed];
\end{tikzpicture}%
}

\newcommand{\given}{\;\middle|\;}

\journal{}
\makeatletter
\def\ps@pprintTitle{%
   \let\@oddhead\@empty
   \let\@evenhead\@empty
   \let\@oddfoot\@empty
   \let\@evenfoot\@oddfoot
}

\begin{document}

\begin{frontmatter}



\title{Incorporating travel behavior regularity into passenger flow forecasting}

\author[label1,label2]{Zhanhong Cheng}
\ead{zhanhong.cheng@mail.mcgill.ca}

\author[label2,label3]{Martin Tr\'epanier}
\ead{mtrepanier@polymtl.ca}

\author[label1,label2]{Lijun Sun\corref{cor1}}
\ead{lijun.sun@mcgill.ca}

\address[label1]{Department of Civil Engineering, McGill University, Montreal, QC H3A 0C3, Canada}
\address[label2]{Interuniversity Research Centre on Enterprise Networks, Logistics and Transportation (CIRRELT)}
\address[label3]{Department of Mathematics and Industrial Engineering, Polytechnique Montreal, Montreal, QC H3T 1J4, Canada}

\cortext[cor1]{Corresponding author. Address: 492-817 Sherbrooke Street West, Macdonald Engineering Building, Montreal, Quebec H3A 0C3, Canada}

\begin{abstract}
Accurate forecasting of passenger flow (i.e., ridership) is critical to the operation of urban metro systems. Previous studies mainly model passenger flow as time series by aggregating individual trips and then perform forecasting based on the values in the past several steps. However, this approach essentially overlooks the fact that passenger flow consists of trips from each individual traveler. For example, a traveler's work trip in the morning can help predict his/her home trip in the evening, while this causal structure cannot be explicitly encoded in standard time series models. In this paper, we propose a new forecasting framework for boarding flow by incorporating the generative mechanism into standard time series models and leveraging the strong regularity rooted in travel behavior. In doing so, we introduce returning flow from previous alighting trips as a new covariate, which captures the causal structure and long-range dependencies in passenger flow data based on travel behavior. We develop the return probability parallelogram (RPP) to summarize the causal relationships and estimate the return flow. The proposed framework is evaluated using real-world passenger flow data, and the results confirm that the returning flow---a single covariate---can substantially and consistently improve various forecasting tasks, including one-step ahead forecasting, multi-step ahead forecasting, and forecasting under special events. And the proposed method is more effective for business-type stations with most passengers come and return within the same day. This study can be extended to other modes of transport, and it also sheds new light on general demand time series forecasting problems, in which causal structure and long-range dependencies are generated by the user behavior.
\end{abstract}


\end{frontmatter}

\section{Introduction} 

Recent years have witnessed the rapid development of metro systems and the continued growth of metro ridership worldwide \citep{metro2018}. As an efficient and high-capacity transportation mode, the metro is playing an ever-important role in shaping future sustainable transportation. Given the growing importance of metro systems, it is critical to have a good understanding of passenger demand patterns to support service operation. A key task is to make accurate and real-time forecasting of passenger demand/ridership, which plays a vital role in a wide range of applications, including service scheduling, crowd management, and disruption response, to name but a few.

Short-term passenger flow forecasting typically focuses on forecasting the passenger flow in the next few minutes to several hours, and has been extensively studied in public transportation research. Most existing studies formulate passenger flow data as time series and follow similar methods as those applied in traffic flow forecasting. For example, statistical time series models have been widely applied to ridership forecasting problems, including auto-regressive integrated moving average (ARIMA) \citep{williams2003modeling, ding2017using,chen2019subway}, exponential smoothing \citep{tan2009aggregation}, and state-space/Kalman filter \citep{stathopoulos2003multivariate, jiao2016three}. Most of these classical time series models are linear by nature; to better characterize the non-linearity in time series data, non-linear versions or ensemble extensions of these models have also been studied \citep[e.g.,][]{jiao2016three, carrese2017dynamic}. Recent research starts regarding the forecasting a supervised machine learning problem. On this track, some representative supervised learning models have been applied, such as support vector machine (SVM) \citep{chen2011use,sun2015novel}, artificial neural network (ANN) \citep{vlahogianni2005optimized, tsai2009neural, li2017forecasting}, random forest \citep{toque2017short}, and recurrent neural network (RNN)/long short-term memory (LSTM) as emerging deep learning approaches \citep{hao2019sequence,liu2019deeppf}. The aforementioned research mainly focuses on modeling a univariate time series for a single metro station. However, the metro system is a network in which stations exhibit strong spatial and temporal correlations/dependencies. To extend the univariate analysis to network-wide passenger flow forecasting, some state-of-the-art models have been proposed to better characterize the complex spatiotemporal patterns and dynamics. For example, \citet{gong2020online} proposed matrix factorization models to estimate passenger flow data for each origin-destination (OD) pair; \citet{li2019tensor} introduced local smoothness prior based on auxiliary information (e.g., flow correlation, network typology, and POI composition) into tensor completion models to forecast passenger flow; \citet{chen2020physical} developed graph convolutional network (GCN) models to capture the complex spatiotemporal dependencies in a metro network. These new machine learning-based models have shown superior performance over traditional time series models, and they are more effective in capturing the complex patterns by incorporating domain knowledge and external features such as weather, event, time of day, and day of week.

In all the studies mentioned above, passenger flow data is generally modeled as an aggregated count time series obtained by counting the number of unique card IDs in smart card transactions. Despite the simplicity and effectiveness of these models, we would argue that the most important characteristic of passenger flow is overlooked due to the aggregation: passenger flow consists of the movement of individuals with strong regularity rooted in their travel behavior. For instance, if a passenger alights at a metro station for work in the morning, he/she will probably depart at the same station when he/she goes home in the evening. If he/she does not travel in the morning, it becomes less likely we will  observe a corresponding return trip. This example clearly shows that past trips should be utilized to predict future demand, and individual travel behavior actually can result in causal structure and long-range dependencies in passenger flow time series data. Some recent studies have shown that travel behavior plays a substantial role in dynamic traffic assignment \citep{cantelmo2019activity} and online demand estimation \citep{cantelmo2019trip}. This effect is particularly true for metro systems where passengers' travel patterns are highly regular \citep{sun2013understanding, goulet2017measuring, zhao2018individual}. Therefore, when developing a passenger flow forecasting model, it is essential to integrate this type of behavior-driven and long-range dependencies in addition to the local input (e.g., the past $n$ steps in the time series).

The goal of this study is to explore the potential of incorporating an additional travel behavior component into the forecasting of passenger flow time series. Specifically, we propose a new scheme to forecast boarding/incoming passenger demand at a station by integrating historical alighting time series at the same station. We define returning passengers as those who finish their first trip at station $s$ and also start their second trip at the same station. In other words, returning passengers refer to the individuals who stay at station $s$ to perform an activity (e.g., home and work). In general, these return trips are not random and often exhibit strong regularity due to the activities performed. This motivates us to forecast the incoming/boarding demand from these ``returning passengers'' using the information on their previous trips. To achieve this, we introduce a new concept of return probability parallelogram (RPP) to better estimate returning flow, and we find that the estimated returning flow highly correlates with the overall boarding demand in a real-world data set. To further quantify the benefits of incorporating this returning flow measure, we evaluate the proposed models for one-step ahead forecasting, multi-step ahead forecasting, and forecasting under special events. Our results show that incorporating returning flow as an additional variable will consistently improve the accuracy of forecasting.

The idea of leveraging trip-level information has been introduced and examined in some recent studies, which predict the alighting flow of a station using the recent boarding flow from other related stations \citep[see e.g.,][]{li2017forecasting,hao2019sequence,liu2019deeppf}. However, the large number of boarding-alighting station pairs makes it difficult to learn an informative model at a trip level, and eventually these studies develop deep neural networks to learn the correlation from the aggregated count data in a purely data-driven approach. Our model, instead, uses the alighting of ``this trip'' to predict the boarding of the ``next trip'', where the alighting and the boarding stations are usually the same \citep{barry2002origin,trepanier2007individual}. We examine this idea on a boarding flow forecasting application, which is more important to service operation and planning. The ``returning flow'' proposed in this paper is solely based on the intrinsic travel regularity of travelers and it does not require external information/knowledge. Our work is closely related to \citet{zhao2018individual}, which proposes a probabilistic model to predict the next trip for an individual based on his/her trip history. However, instead of predicting individual trips, our primary goal is to forecast the overall passenger flow to support the decision making in service operation. In doing so, we estimate the returning flow in an aggregated approach; therefore, the framework does not require individual-based data sets that are confidential and sensitive for privacy reasons. The main contribution of this work is summarized as follows.
\begin{itemize}
      \item We define returning flow to characterize the causal structure and long-range dependencies in passenger flow data, which are essentially overlooked in previous time series-based studies.
      \item We integrate returning flow as an additional covariate into standard time series models, and the proposed behavior-integrated model shows consistently improved performance in our case studies based on a real-world data set.
      \item Our model also provides a new approach to forecast passenger flows under special events.
\end{itemize}

To the best of our knowledge, this is the first research that incorporates a travel behavior component into the longstanding passenger flow forecasting problem. The remainder of the paper is organized as follows. Section~\ref{sec:model} introduces the concept of returning flow and return probability parallelogram as the tool to integrate travel behavior regularity into the passenger flow forecasting framework. In section~\ref{sec:experiment}, we develop case studies based on real-world smart card data and demonstrate the effectiveness of the proposed models in different scenarios. Finally, section~\ref{sec:conclusion} concludes our research and discusses future work.

\section{Methodology}\label{sec:model}


In this section, we introduce returning flow and the return probability parallelogram as two fundamental building blocks in the behavior-based boarding flow forecasting framework. The proposed forecasting models are constructed by integrating returning flow as a new feature/covariate into traditional time series forecasting models. We start with a brief description of the passenger flow forecasting problem.


\subsection{Problem description}

Suppose that in a metro system we have access to all smart card transactions, i.e., we know the anonymous ID of passengers, the time and the locations/stations of both boarding (tapping-in) and alighting (tapping-out) for each trip. In this case, a station $s$ will generate two passenger flow time series: the alighting/arriving flow for passengers with station $s$ as their destination, and the boarding/incoming flow for passengers who start their trips from station $s$. We denote by $y^s_t$ and $m^s_t$ the boarding flow and the alighting flow at station $s$ in time interval $t$, respectively.

We focus on the case of forecasting the boarding flow $y^s_t$. Given some recent observations $y^s_{1},\ldots,y^s_{t-1}, y^s_{t}$, our goal is to predict the values of $y^s_{t+1}, y^s_{t+2}, \ldots, y^s_{t+L}$ in the next $L$ time steps/intervals. This is a standard time series analysis problem on which traditional statistical models such as ARIMA can be applied. In this paper, we aim to achieve better forecasting results over the traditional models by integrating additional information of the behavioral regularity of passengers associated with alighting flow $m^s_t$.

\subsection{Returning flow}

We begin our model by introducing the concept of the ``returning flow''. To facilitate model development, we divide all the passengers associated with stations $s$ (both boarding and alighting) into two groups (see Figure~\ref{fig:example}):
\begin{description}
    \item[(G1)] Passengers who alight at station $s$;
    \item[(G2)] Passengers who board at station $s$ without a previous trip alighting at $s$.
\end{description}

\begin{figure}[!ht]
      \centering
      \small
      \begin{tabular}{cc|l|c c c c c| c c c c}
      \hline
      \multicolumn{2}{c|}{Group} &       &        $t-h$   &   $\ldots$    & $t-2$ & $t-1$ & $t$ & $t+1$ & $t+2$ & $\ldots$ & $t+L$ \\
      \hline
      \multirow{8}{*}{G1} & \multirow{6}{*}{A} &        &\tikzmark{a1}{$\circ$} &       &\tikzmark{b1}{$\bullet$} &       &       &       &       &       &  \\
      &     &     &     &  \tikzmark{a2}{$\circ$} &       &       &  \tikzmark{b2}{$\bullet$}       &       & &        &  \\
      &     &     &       &       & &  &     \tikzmark{a6}{$\circ$}  &       &      \tikzmark{b6}{$\bullet$}    &   &  \\
      &     &     &       &       &       &       & \tikzmark{a3}{$\circ$} &       &       &       & \tikzmark{b3}{$\bullet$} \\
      &     &     &       &       &       & \tikzmark{a4}{$\circ$} &       & \tikzmark{b4}{$\bullet$} &       &       &  \\
      &     &     &       &  &\tikzmark{a5}{$\circ$} &             &       &       & \tikzmark{b5}{$\bullet$} &       &  \\
      \cline{2-2}\cline{3-12}
      & \multirow{2}{*}{B}     &        &       &        &   $\circ$ &    &       &       &       &       &  \\
      &     &        &       &       &       & $\circ$ &       &       &       &       &  \\
      \hline
      &     & sum $\circ$ in G1 & $m^s_{t-h}$ &    $\ldots$    &  $m^s_{t-2}$      &  $m^s_{t-1}$     &  $m^s_{t}$      &       &       &       &  \\
      \hline
      \multicolumn{2}{c|}{\multirow{2}{*}{G2}} &        &       &       &       &       & $\bullet$ &       &       &       &  \\
      &     &    &       &       &       &       &       &    $\bullet$    & &       &  \\
      \hline
      &       & sum $\bullet$ in G1+G2  & $y^s_{t-h}$ &     $\ldots$   &   $y^s_{t-2}$    & $y^s_{t-1}$      & $y^s_{t}$      &  $\hat{y}^s_{t+1}$     &       &       &  \\
      \hline
      \end{tabular}%
      \link{a1}{b1}
      \link{a2}{b2}
      \dashlink{a3}{b3}
      \dashlink{a4}{b4}
      \dashlink{a5}{b5}
      \dashlink{a6}{b6}
      \vspace{1em}

       $\circ$ represents alighting, $\bullet$ represents boarding, \\
      $\longrightarrow$ represents observed trip chain, $\dashrightarrow$ represents trip chain to be predicted.
      \caption{Illustration of two passenger groups (G1/2) and the boarding demand forecasting problem at station $s$.}
      \label{fig:example}
\end{figure}

\begin{figure}
    \centering
    \includegraphics[scale=0.8]{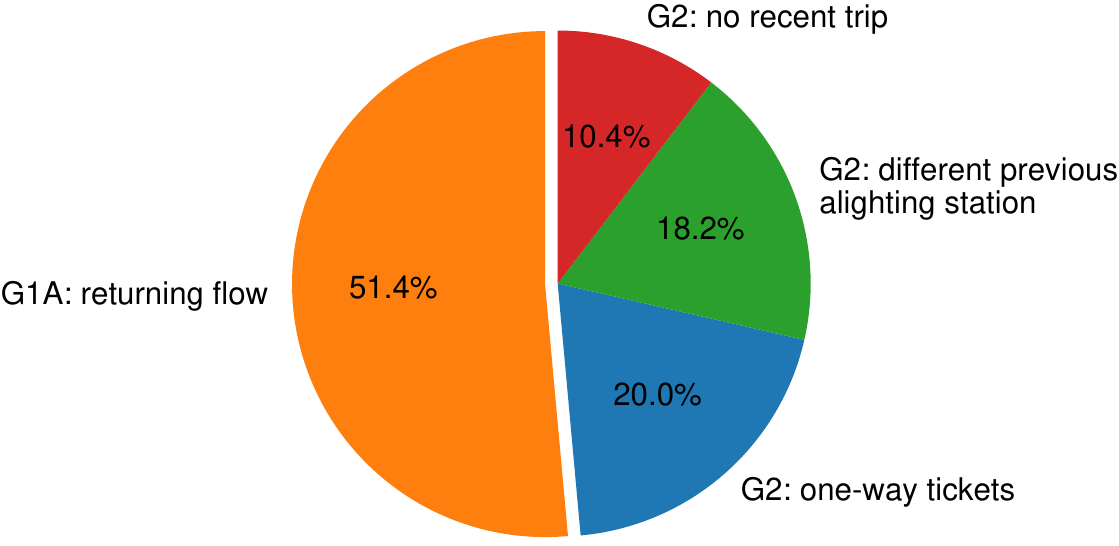}
    \caption{ The composition of the boarding flow in Guangzhou metro. Based on the smart card data from July 21 to 28, 2017.}
    \label{fig:pie}
\end{figure}

With this definition, we can model the total boarding flow $y^s_t$ by combining the boarding flow in the two groups. The passengers in G1 can be further separated into two subgroups given if they have their next trip originating from station $s$ within a certain time window. We define the subgroup with a following trip as G1A and the other as G1B. Thus, G1A actually consists of those passengers who conduct certain activities (e.g., home/work) around station $s$. We define ``returning flow'' at time $t$ as the number of people in G1 who will finish their activities and start their return trips at time $t$ by station $s$, denoted by $r^s_{t}$. In fact, these chained trips (with departing station identical to the alighting station of a previous trip) make up a substantial proportion of all trips. As shown in Figure~\ref{fig:pie}, for the Guangzhou metro in our case study, the returning flow accounts for over $50\%$ of all boarding demand. We thus hypothesize that having the ``returning flow'' as an additional variable will enhance the forecasting of $y^s_{t+1}$. We refer to the forecasting model with $r^s_{t+1}$ as a covariate as M2:
\begin{equation}
\text{M2}: \ \ \hat{y}^s_{t+1} = f\left(y^s_{1:t}, r^s_{t+1}\right).
\end{equation}

It should be noted that we do not have access to $r^s_{t+1}$ (i.e., those dashed arrows in Figure~\ref{fig:example}), as the returning flow in G1 is only observed up to time $t$ (i.e., those solid arrows in Figure~\ref{fig:example}). Therefore, in practice, we need to first estimate $\hat{r}^s_{t+1}$ and then use it as a proxy for $r^s_{t+1}$ in M2. On the other hand, a possible alternative is to use $r^s_t$---which we have access in real-time---instead of $\hat{r}^s_{t+1}$ as the covariate. We define this alternative model as M1 and use it as a baseline model:
\begin{equation}
\text{M1}: \ \ \hat{y}^s_{t+1} = f\left(y^s_{1:t}, r^s_{t}\right).
\end{equation}
We also consider a standard time series model without any additional variables as a baseline (M0):
\begin{equation}
\text{M0}: \ \ \hat{y}^s_{t+1} = f\left(y^s_{1:t}\right).
\end{equation}

Note the returning flow (G1A) in this paper does not cover the G2 part of the boarding flow (Figure~\ref{fig:pie}). Because it is hard to forecast the one-way tickets or standalone trips in G2 by data solely from a metro system. However, a metro ride can be one trip in an activity chain or even one mode in a multi-modal trip; it is possible to infer the G2 part if we have complete trip chain information supported by other data sources, which will be greatly helpful for the boarding flow forecasting. Like in most cases, we have no complete trip/activity chain information. We thus establish the returning flow concept upon consecutive metro trips with the first destination and the next origin overlapped. This is a specific subset of activity chains. Luckily, we can obtain a very accurate estimation for the future returning flow (G1A) and it already takes a substantial part of the total boarding flow; using the returning flow as a covariate is still beneficial for the boarding flow forecasting.

\subsection{Return probability parallelogram (RPP)}\label{sec:rpp}
In this subsection, we propose a method to estimate the returning flow. We only consider the returning flow within a time window $H$ when we define G1A. For the current and past returning flow, $r^s_t$ can be readily obtained by
\begin{equation}\label{eq:return_t}
      r_t^s = \sum_{t_{a}=t-H}^{t-1}r_{t_{a},t},
\end{equation}
where $r_{t_{a},t}$ is the number of passengers that come (alight) at station $s$ at $t_{a}$ and return (board) at $t$. Using Eq.~\eqref{eq:return_t}, $r_t^s$ can be obtained in real-time and used in M1.

However, M2 requires a returning flow in the future that cannot be accessed by Eq.~\eqref{eq:return_t}. Therefore, we propose a method to estimate $\hat{r}_{t+1}^s$ based on the returning flow generalization mechanism. Our fundamental assumption is that there exists a universal distribution $p^s\left(\tau_{\text{boarding}} \given \tau_{\text{alighting}} \right)$ characterizing the conditional probability that a passenger in G1 who alights at time $\tau_{\text{alighting}}$ will start his/her returning trip at time $\tau_{\text{boarding}}$. Note that we define $p^s$ on the whole group G1, so the subgroup G1B is also modeled in this distribution. For the passengers who alight at time $t_a$ we have:
\begin{equation}
    \sum\limits_{t=t_a+1}^{t_a+H}p^s\left(\tau_{\text{boarding}}=t \given \tau_{\text{alighting}}=t_a \right) + p^s\left(\tau_{\text{boarding}}=\text{NA} \given \tau_{\text{alighting}}=t_a \right) =1,
\end{equation}
in which the term $p^s\left(\tau_{\text{boarding}}=\text{NA} \given \tau_{\text{alighting}}=t_a \right)$ represents the conditional probability of an arriving passenger does not return within the time window $H$ (i.e., subgroup G1B).

If the conditional distribution $p^s\left(\tau_{\text{boarding}}\given \tau_{\text{alighting}}\right)$ is available for all $\tau_{\text{alighting}}$, we can estimate the expectation of the returning flow $r^s_{t+1}$ at time $t+1$ by:
\begin{equation}\label{equ:return_tplus1}
\hat{r}^{s}_{t+1}=\sum_{h=1}^{H}{m^{s}_{t-h+1} {p}^{s}\left(\tau_{\text{boarding}}=t+1 \given \tau_{\text{alighting}}=t-h+1\right)}.
\end{equation}

It is important to note that the estimation of $\hat{r}^s_{t+1}$ using Eq.~\eqref{equ:return_tplus1} is very different from predicting $\hat{r}^s_{t+1}$ using a time series model based on past observations. This is because a simple time series model such as ARIMA cannot characterize the unique generative mechanisms (e.g., come-and-return) and the corresponding long-range dependencies/causal structure provided by these mechanisms in the passenger flow data.

\begin{figure}[!ht]
    \begin{center}
    \includegraphics[]{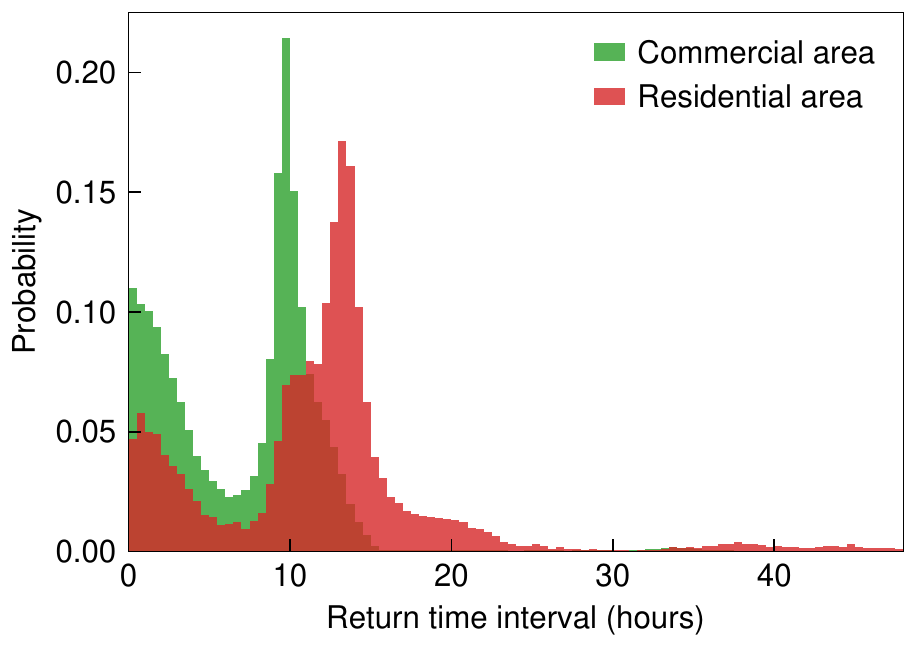}
    \caption{The density histogram of the return time interval ($\tau_{\text{boarding}} - \tau_{\text{alighting}}$) of two example stations.}
    \label{fig:duration}
    \end{center}
\end{figure}

The time window length $H$ is an additional parameter to be determined before applying Eq.~\eqref{equ:return_tplus1}. To choose an appropriate $H$, we quantify the inter-trip time/duration $\left(\tau_{\text{boarding}}-\tau_{\text{alighting}}\right)$ for all those passengers with both the alighting trip and the next boarding trip at the same station. We conduct this analysis on the Guangzhou metro data set. Figure~\ref{fig:duration} shows the distribution of the inter-trip time of two representative stations in a commercial area and a residential area, respectively. The distribution is obtained by aggregating all alighting records on a typical Monday, and we track the returning flow within 48 hours after the alighting. The return time intervals in both stations are characterized by a bi-modal pattern. The first peak (less than 3 hours) corresponds to certain short-duration activities (e.g. dining and shopping). The longer peaks largely correspond to ``work'' activities (9-12 hours) in the commercial area and ``home'' activities (10-16 hours) in the residential area, respectively. Relatively few activities take around 6 hours, and thus the return time intervals in both stations exhibit a ``U'' shape pattern. More importantly, as we can see, almost all of the return trips start within a 24-hour window after finishing previous trips. Therefore, for simplicity, we only take the alighting flow within the past 24 hours into account when estimating $\hat{r}^s_{t+1}$.

Having determined $H$, the next step is to obtain a good estimate of the conditional probability distribution $p^s\left(\tau_{\text{boarding}} \given \tau_{\text{alighting}} \right)$. However, the current formulation involves a set of conditional probabilities for each value of $\tau_{\text{alighting}}$, making it difficult to estimate. For simplicity, we assume that the conditional distributions are universal across different days:
\begin{equation}
\begin{split}
    p^s\left(\tau_{\text{boarding}}=t_b \given \tau_{\text{alighting}}=t_a \right)&  =  p^s_0 \left( \text{future window} (t_b)  \given \text{window of day} (t_a) \right)\\
    & =  p^s_0 \left( \text{window of day} (t_a)+t_b-t_a  \given \text{window of day} (t_a) \right),
\end{split}
\end{equation}
where we refer to the reduced distribution $p^s_0$ as the return probability parallelogram (RPP). As the new conditional distribution $p^s_0$ is defined given the time of day of $t_a$, we can estimate it using historical trip data of passengers in group G1A (i.e., the solid arrows in Figure~\ref{fig:example}). Denote $r_{t_{a},t_{b}}^{s}$ to be the number of passengers that come (alight) at station $s$ at $t_{a}$ and return (board) at $t_{b}$. For a time window $w$ of day, $p^s_0$ can be estimated by
\begin{equation}
  p_{0}^{s}(w+h|w) =
  \frac{
     \sum_{\substack{\text{window of day}(t_a) =w\\
        t_{b}-t_{a} = h}}
     {r_{t_{a},t_{b}}^{s}}}
     {\sum_{\text{window of day}(t_{a})=w}{m_{t_{a}}^{s}}}\quad (h=1, 2, \cdots, H).
\end{equation}

\begin{figure}[!ht]
\begin{center}
\includegraphics[]{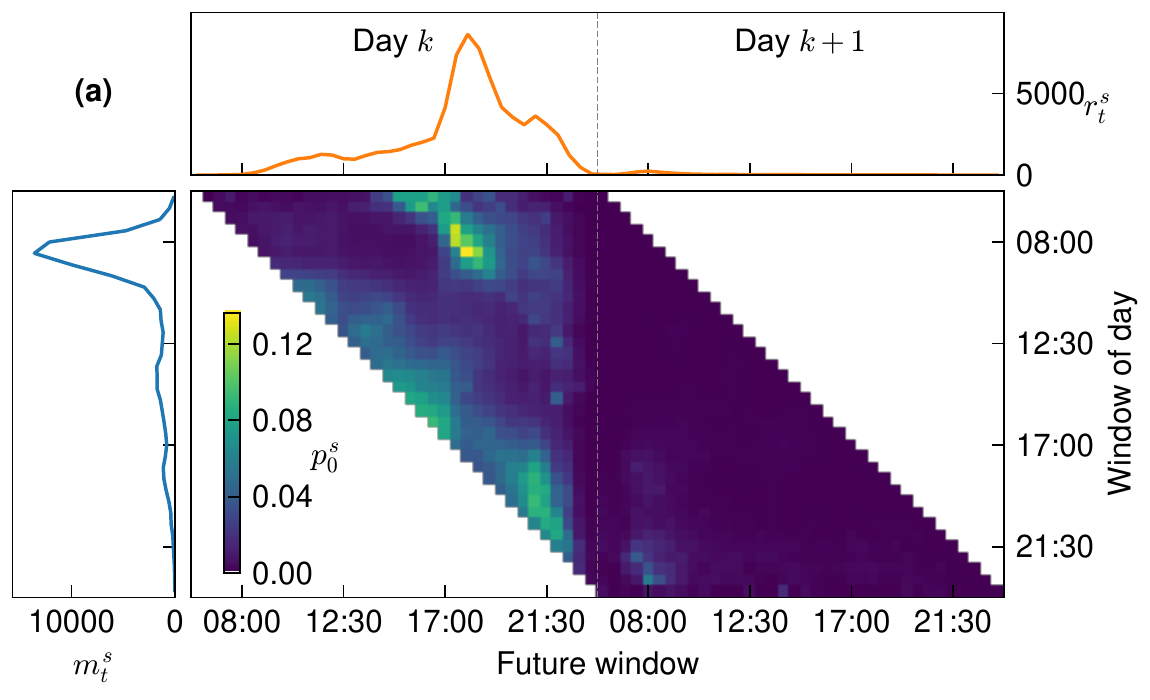}
\includegraphics[]{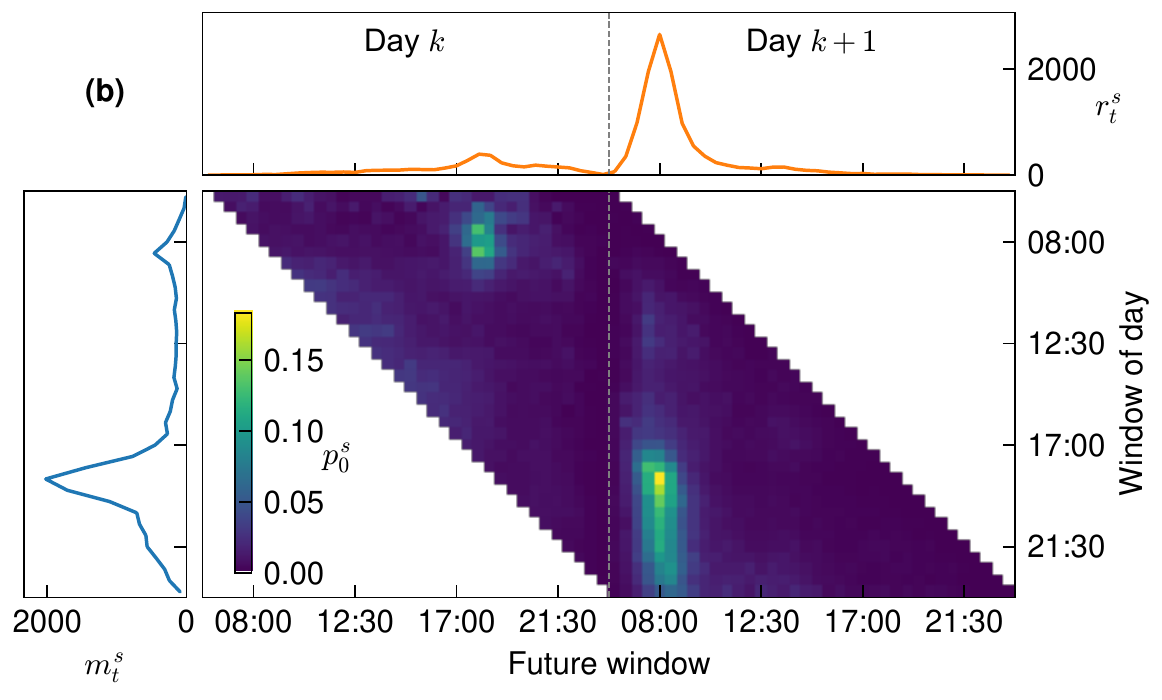}
\caption{The return probability parallelogram (RPP), the alighting flow $m^{s}_{t}$, and the returning flow $r^{s}_{t}$ for two representative stations: (a) A typical station in commercial areas. (b) A typical station in residential areas.}
\label{fig:rpp}
\end{center}
\end{figure}

We use Figure~\ref{fig:rpp} to illustrate the idea of RPP. Panel (a) and (b) show two sets of conditional distributions for a commercial area and a residential area, respectively, in Guangzhou metro. The resolution for time slot is set to half an hour, and the range is from 6:00 to 24:00 (operation time of the metro system). Note that in this parallelogram representation we concatenate the 24:00 of day $k$ and the 6:00 of day $k+1$ on the horizontal axis. There are two blank triangles in this diagram: the one on the left corresponds to the $t_b\le t_a$, where the distribution is not defined; the one on the right corresponds to the conditional probability with $t_b> t_a+H$ ($H=48$ for 24 hours), which is also ignored for simplicity. It should be noted that in RPP the sum of each row is less than 1, as it does not include the passengers in G1B (with no returning trips, i.e., $\tau_{\text{boarding}}=\text{NA}$). With this formulation, we can replace ${p}^{s}\left(\tau_{\text{boarding}}=t+1 \given \tau_{\text{alighting}}=t-h+1\right)$ in Eq.~\eqref{equ:return_tplus1} by the corresponding conditional probability in RPP.

As shown in Figure~\ref{fig:rpp}, it is obvious that different stations exhibit different RPP patterns. For example, for the commercial station in Figure~\ref{fig:rpp}(a), most trips arrive (alight) in the morning and return in the evening on the same day, which essentially captures work activities. It is very rare to see returning trips on the next day. As for the station in a residential area in Figure~\ref{fig:rpp}(b), on the contrary, we can see that the distribution mainly characterizes home activities, where alighting flow generally peaks in the evening and the returning flow concentrates in the morning of the next day. The RPP representation demonstrated in Figure~\ref{fig:rpp} further suggests that the unique come-and-return dynamics for a station should be considered in passenger flow forecasting applications.

\section{Experiments}\label{sec:experiment}

In this section, we conduct numerical experiments to evaluate the effectiveness of the proposed behavior-integrated models. We choose the standard SARIMA model as the core model for time series forecasting (M0). On top of this model, we create two regression with SARIMA error models---M1 and M2---by simply incorporating the observed $r^s_t$ and the estimated $\hat{r}^{s}_{t+1}$ as additional covariates, respectively. We evaluate the performance of these models in three scenarios: 1) one-step ahead forecasting, 2) multi-step ahead forecasting, and 3) forecasting under special events. Besides, we also test using Support Vector Regression (SVR) and Multi-Layer Perceptron (MLP) as M0 in \ref{appendix:A} and observe consistent results with the SARIMA.

\subsection{ARIMA model}

We choose seasonal ARIMA as the main baseline model---M0. ARIMA is a well-established time series forecasting model which has been widely used in traffic/passenger flow forecasting \citep{williams2003modeling,ding2017using,chen2019subway}. Considering the strong periodicity from day to day, we apply Seasonal ARIMA (SARIMA) to model passenger flow. Here we give a brief introduction of the SARIMA model, and we refer readers to \citet{hyndman2018forecasting} for a comprehensive review of time serise models. A SARIMA model is usually denoted by ARIMA$(p,d,q)(P,D,Q)[m]$, where $p$, $d$, and $q$ represent the order of autoregressive, differencing, and moving-average; $P$, $D$, and $Q$ are the order of autoregressive, differencing, and moving-average for the seasonal part; and $m$ is the number of period in each season. For a time series $y_{1}, \ldots, y_{T}$, the ARIMA$(p,d,q)(P,D,Q)[m]$ model (M0) takes the form
\begin{equation}\label{eq:arima}
\Phi(B)\left(1-B^{m}\right)^{D}\phi(B)(1-B)^{d}
y_{t}=
\theta(B)\Theta(B)e_{t},
\end{equation}
where $B$ is the backshift notation defined by $B^{a}y_{t}=y_{t-a}$,
$\Phi(B) = \left(1-\Phi_{1} B^{m}-\cdots-\Phi_{P} B^{P \times m}\right)$,
$\phi(B) = \left(1-\phi_{1} B-\cdots-\phi_{p} B^{p}\right)$,
$\theta(B) = \left(1+\theta_{1} B+\cdots+\theta_{q} B^{q}\right)$, and
$\Theta(B) = \left(1+\Theta_{1} B^{m}+\cdots+\Theta_{Q} B^{Q\times m}\right)$; $\Phi_{i}$, $\Theta_{i}$, $\phi_{i}$, and $\theta_{i}$ are ARIMA coefficients to be estimated; $e_{t}$ is an error assumed to follow a white noise process (i.e., zero mean and iid).

When incorporating the returning flow $r_{1}, \ldots, r_{T}$, as a covariate, the forecasting model (M2) becomes
\begin{equation}\label{eq:regression}
   \begin{gathered}
      y_{t} = \beta r_{t} + \eta_{t},\\
      \Phi(B)\left(1-B^{m}\right)^{D}\phi(B)(1-B)^{d}\eta_{t}=
      \theta(B)\Theta(B)e_{t}.
   \end{gathered}
\end{equation}
Where $\beta$ is the regression coefficient, $\eta_{t}$ is a regression error term that follows the ARIMA procedure. Note the regression coefficient and the ARIMA coefficients are estimated in one step, rather than estimated separately. All ARIMA models in this study are estimated using the \textbf{forecast} package for R \citep{R2020}.

\subsection{Model selection and evaluation}

We apply the same order of SARIMA to the demand time series for all the stations. The seasonal frequency is set to $m=36$ (i.e., daily, from 6:00 to 24:00, as we use half an hour as the temporal resolution). For most stations, after a seasonal differencing, the Augmented Dickey-Fuller (ADF) test \citep{dickey1979distribution} indicates no further differencing is required to make the time series stationary, we thus set $D=1$ and $d=0$. We search over possible models and finally select ARIMA$(2,0,1)(1,1,0)[36]$ as the baseline M0, which is shown to be appropriate for most stations. Indeed, we can achieve better forecasting results by designing station-specific models with different orders. However, as our goal is to evaluate the effect of using the returning flow as a covariate, we still select a universal model for all stations for simplicity.

We use the root mean square error (RMSE) and the symmetric mean absolute percentage error (SMAPE) to evaluate model accuracy:
\begin{equation}\label{eq:rmse}
      \text{RMSE} = \sqrt{\frac{1}{N} \sum_{t=1}^{N}(y_{t}^{s} - \hat{y}_{t}^{s})^{2}},
\end{equation}
\begin{equation}\label{eq:smape}
      \text{SMAPE} = \frac{2}{N} \sum_{t=1}^{N}{\frac{|y_{t}^{s} - \hat{y}_{t}^{s}|}{|y_{t}^{s}| + |\hat{y}_{t}^{s}|}}\times 100 (\%),
\end{equation}
where $y_{t}^{s}$ and $\hat{y}_{t}^{s}$ are the real boarding flow and the predicted boarding flow, respectively. In addition to RMSE and SMAPE, we also use the Akaike information criterion (AIC) \citep{akaike1998information} to measures the trade-off between the goodness of fit and the complexity of a model. A smaller AIC suggests a better model.

\subsection{Data}\label{sec:data}
We use the passenger flow data retrieved from Guangzhou metro in China as a case study. The smart card data set covers 159 stations from July 24 to September 8 in 2017. Note that the data on weekends are not included in our analysis (i.e., we concatenate Friday with the next Monday), as the RPP has different patterns in weekends. We divide the whole data set into three parts:
\begin{description}
    \item[(D1)] July 24 to August 4 (two weeks): estimate the RPP $p^s_0$ for a station $s$ (for M2 only);
    \item[(D2)] August 7 to August 25 (three weeks): estimate model parameters for all the three SARIMA models (training set);
    \item[(D3)] August 28 to September 8 (two weeks): evaluate model performance (test set).
\end{description}

\begin{figure}[!ht]
\begin{center}
\includegraphics[]{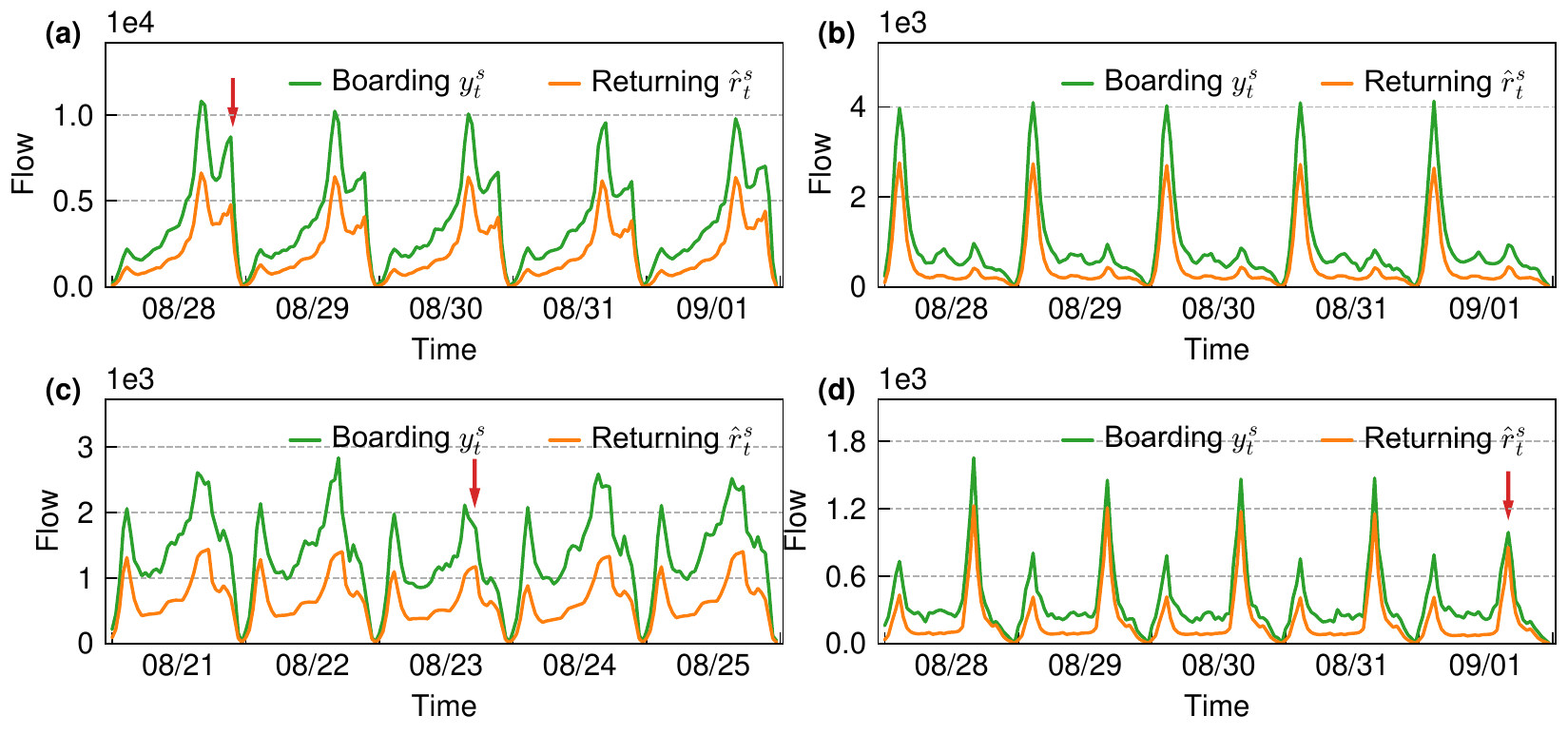}
\caption{The boarding flow and the estimated returning flow. As marked by red arrows, the returning flow reflects the irregular increases/drops of the boarding flow: (a) Tiyu Xilu station, (b) Luoxi station, (c) Changshou Lu station, and (d) Huijiang station. The green curves represent demand time series $y^s_t$ and the orange curves correpsond to the estimated returning flow time series $\hat{r}^{s}_t$.}\label{fig:flows}
\end{center}
\end{figure}

After estimating RPP from D1, we compute $\hat{r}^{s}_{t+1}$ on data sets D2 and D3 following Eq.~\eqref{equ:return_tplus1}. Before estimating the SARIMA models, we first empirically examine the relationship between the time series of returning flow $\hat{r}^{s}_{t}$ and the time series of incoming demand $y^s_{t}$. Figure~\ref{fig:flows} shows the demand time series $y^s_t$ and the estimated returning flow time series $\hat{r}^{s}_t$ on data set D2/D3 for four representative stations. Station (a) and (d) are commercial areas, where the boarding flows concentrate in the afternoon and evening. Station (b) is a residential area that have an extremely high morning peak. Station (c) has high boarding flow in both the morning and afternoon peaks. We can see that the estimated returning flow $\hat{r}^{s}_t$ matches $y^s_t$ for all types of stations very well. Notably, the returning flow $\hat{r}^{s}_t$ makes up a large proportion of the total boarding demand, and it can correctly characterize the temporal dynamics in $y^s_t$. More importantly, as marked by the red arrows in panels (a), (c) and (d), the returning flow can even reproduce some irregular increases/drops (i.e., anomalies) of the boarding flow, which are very difficult to capture using conventional time series models with $y^s_t$ alone. 
\subsection{One-step ahead forecasting}\label{sec:short term}

We use data set D2 to estimate model parameters and apply the model to D3 for evaluation. Table~\ref{tab:result} shows the results of one-step ahead forecasting for the four stations in Figure~\ref{fig:flows}. Compared with M0, M2 consistently reduces the RMSE and SMAPE of both training and test sets of the four stations by incorporating $\hat{r}_{t+1}$. Meanwhile, M2 is also superior with larger log-likelihood and lower AIC. However, with the observed $r^s_t$ as input, M1 performs almost the same with M0. This might be due to fact that $r^s_t$ correlates highly with $y^s_t$ (the observation at the last step), since the returning flow covers a considerable proportion of the overall boarding flow. Thus the amount of additional information brought by this term is rather marginal. While on the contrary, $\hat{r}^s_{t+1}$ estimated externally by combining RPP and the alighting time series $m^s_t$ actually encodes the generative mechanisms and long-range dependencies, and thus M2 produces much better forecasting results.

\begin{table}[!ht]
  \centering \small
  \caption{The one-step boarding flow forecasting of four stations.}
      \begin{tabular}{llrrrrrr}
            \toprule
            Stations & Model &  \makecell{RMSE \\ (train)} & \makecell{RMSE \\ (test)} & \makecell{SMAPE \\ (train)} & \makecell{SMAPE \\ (test)} & \makecell{Log-likelihood} & AIC \\
      \midrule
        \multirow{3}[2]{*}{(a) Tiyu Xilu} & M0    & 398.09 & 363.06 & 11.89\% & 11.63\% & -3747.19 & 7504.37 \\
        & M1    & 396.95 & 362.00 & 12.36\% & 12.22\% & -3745.76 & 7503.52 \\
        & M2    & \textbf{372.94} & \textbf{319.60} & \textbf{10.59\%} & \textbf{9.29\%} & \textbf{-3713.9} & \textbf{7439.79} \\
        \midrule
        \multirow{3}[2]{*}{(b) Luoxi} & M0    & 64.06 & 71.61 & 9.51\% & 9.93\% & -2830.25 & 5670.49 \\
        & M1    & 64.06 & 71.61 & 9.51\% & 9.93\% & -2830.25 & 5672.49 \\
        & M2    & \textbf{63.92} & \textbf{71.37} & \textbf{9.48\%} & \textbf{9.89\%} & \textbf{-2829.25} & \textbf{5670.48} \\
        \midrule
        \multirow{3}[2]{*}{(c) Changshou Lu} & M0    & 93.78 & 94.34 & 10.36\% & 11.88\% & -3018.66 & 6047.31 \\
        & M1    & 93.64 & 95.16 & 10.31\% & 11.99\% & -3017.84 & 6047.67 \\
        & M2    & \textbf{90.02} & \textbf{92.74} & \textbf{9.39\%} & \textbf{10.83\%} & \textbf{-2998.07} & \textbf{6008.15} \\
        \midrule
        \multirow{3}[2]{*}{(d) Huijiang} & M0    & 37.32 & 47.37 & 14.05\% & 14.27\% & -2557.94 & 5125.88 \\
        & M1    & 37.29 & 47.29 & 14.05\% & 14.25\% & -2557.61 & 5127.21 \\
        & M2    & \textbf{36.67} & \textbf{38.61} & \textbf{13.49\%} & \textbf{13.91\%} & \textbf{-2549.47} & \textbf{5110.93} \\
        \bottomrule
        \end{tabular}%
  \label{tab:result}
\end{table}%

To further evaluate whether the improvement of M2 is statistically significant, we apply paired t-test to compare the absolute forecast errors on the test set D3. For each station, denote the forecast error of model M to be a random variable $\varepsilon_\mathrm{M} = \hat{y} - y$. When comparing M2 and M0, the null hypothesis $\mathrm{H}_0: \mu\left(|\varepsilon_{\mathrm{M2}}| -|\varepsilon_{\mathrm{M0}}|\right)=0$ means no significant difference between the absolute forecast error of M2 and M0. We use the lower-tailed alternative hypothesis $\mathrm{H}_a: \mu\left(|\varepsilon_{\mathrm{M2}}| -|\varepsilon_{\mathrm{M0}}|\right) < 0$, which means the absolute forecast error of M2 is smaller  than M0. We also compare M2 with M1 in the same way. Based on the p-values in Table~\ref{tab:ttest_arima}, we reject $\mathrm{H}_0$ for stations (a)(c)(d). Therefore, M2 indeed improves the forecast for stations (a)(c)(d). Although M2 also reduces the RMSE and SMAPE for station (b), Table~\ref{tab:ttest_arima} shows the improvement is not significant at the 0.05 level.

\begin{table}[htbp]
      \centering\small
      \caption{Paired t-test p-values for absolute forecast errors.}
      \begin{threeparttable}
        \begin{tabular}{lrrrr}
        \toprule
              & \multicolumn{1}{l}{(a) Tiyu Xilu} & \multicolumn{1}{l}{(b) Luoxi} & \multicolumn{1}{l}{(c) Changshou Lu} & \multicolumn{1}{l}{(d) Huijiang} \\
        \midrule
        \makecell{$\mathrm{H}_0: \mu\left(|\varepsilon_{\mathrm{M2}}| -|\varepsilon_{\mathrm{M0}}|\right) = 0$\\ $\mathrm{H}_a: \mu\left(|\varepsilon_{\mathrm{M2}}| -|\varepsilon_{\mathrm{M0}}|\right) < 0$} & $<0.001^{*}$ & 0.065 & $0.016^*$ & $0.003^*$ \\
        \midrule
        \makecell{$\mathrm{H}_0: \mu\left(|\varepsilon_{\mathrm{M2}}| -|\varepsilon_{\mathrm{M1}}|\right) = 0$\\ $\mathrm{H}_a: \mu\left(|\varepsilon_{\mathrm{M2}}| -|\varepsilon_{\mathrm{M1}}|\right) < 0$} & $<0.001^{*}$ & 0.061 & $0.013^*$ & $0.003^*$ \\
        \bottomrule
        \end{tabular}%
        \begin{tablenotes}
      \item $^*$ Significant at 0.05 level.
        \end{tablenotes}
      \end{threeparttable}
      \label{tab:ttest_arima}%
\end{table}%

The results in Table~\ref{tab:result} and Table~\ref{tab:ttest_arima} indeed show that M2 gives improved accuracy; however, it should be also noted that the improvement varies across different stations. To further explore this variation, we cluster the 159 stations based on their RPPs. In doing so, we transform each RPP into a vector of $36 \times 36=1296$ and perform hierarchical clustering using the Euclidean distance between paired vectors; the distances between clusters are calculated by the Ward's method \citep{ward1963hierarchical}. In the meanwhile, we measure the effect of $\hat{r}_{t+1}^s$ by the difference in SMAPE:
\begin{equation}
D^{s} = \text{SMAPE}^{s}_{\text{M2}} - \text{SMAPE}^{s}_{\text{M0}},
\end{equation}
where $\text{SMAPE}^{s}_{\text{M2}}$ and $\text{SMAPE}^{s}_{\text{M2}}$ are SMAPE values of M2 and M0, respectively, on the test data set D3. A negative $D^{s}$ means M2 improves the forecasting accuracy.

\begin{figure}[!ht]
      \begin{center}
      \includegraphics[]{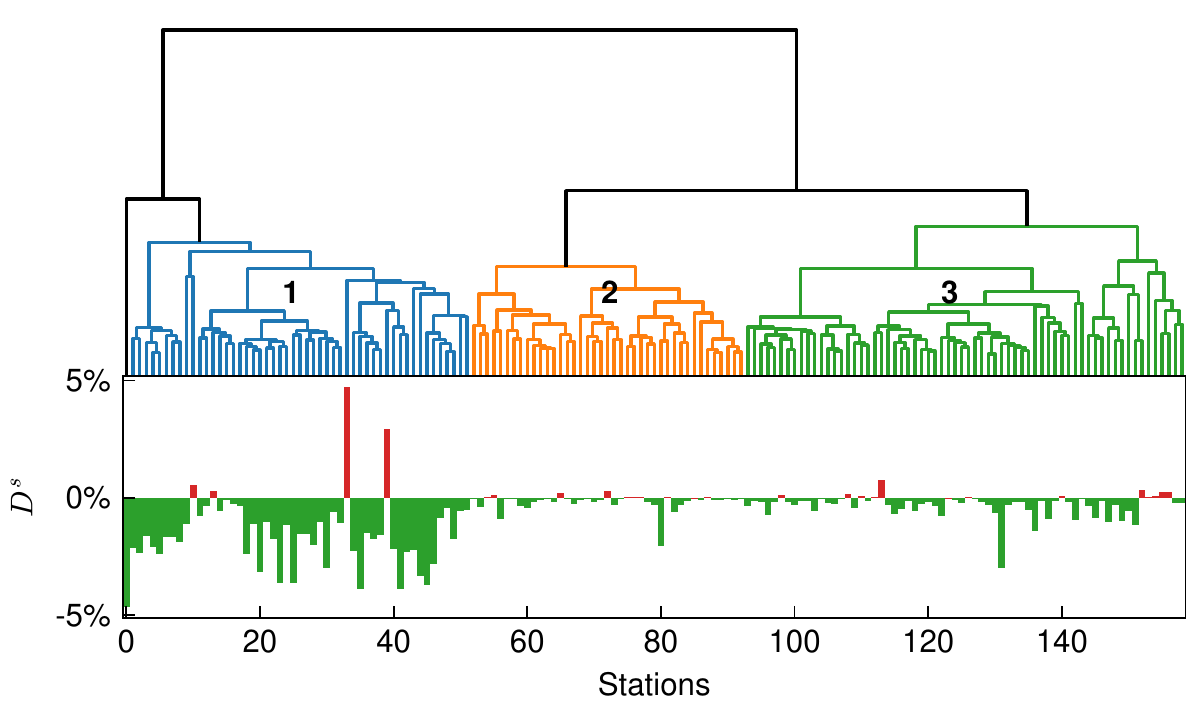}
      \caption{Top: the dendrogram for the hierarchical clustering based on RPP. Bottom: the test set SMAPE differences between M2 and M0; green and negative values means using $\hat{r}_{t+1}^s$ improves the boarding flow forecast in the test set.}\label{fig:cluster}
      \end{center}
\end{figure}

\begin{figure}[!ht]
      \begin{center}
      \includegraphics[]{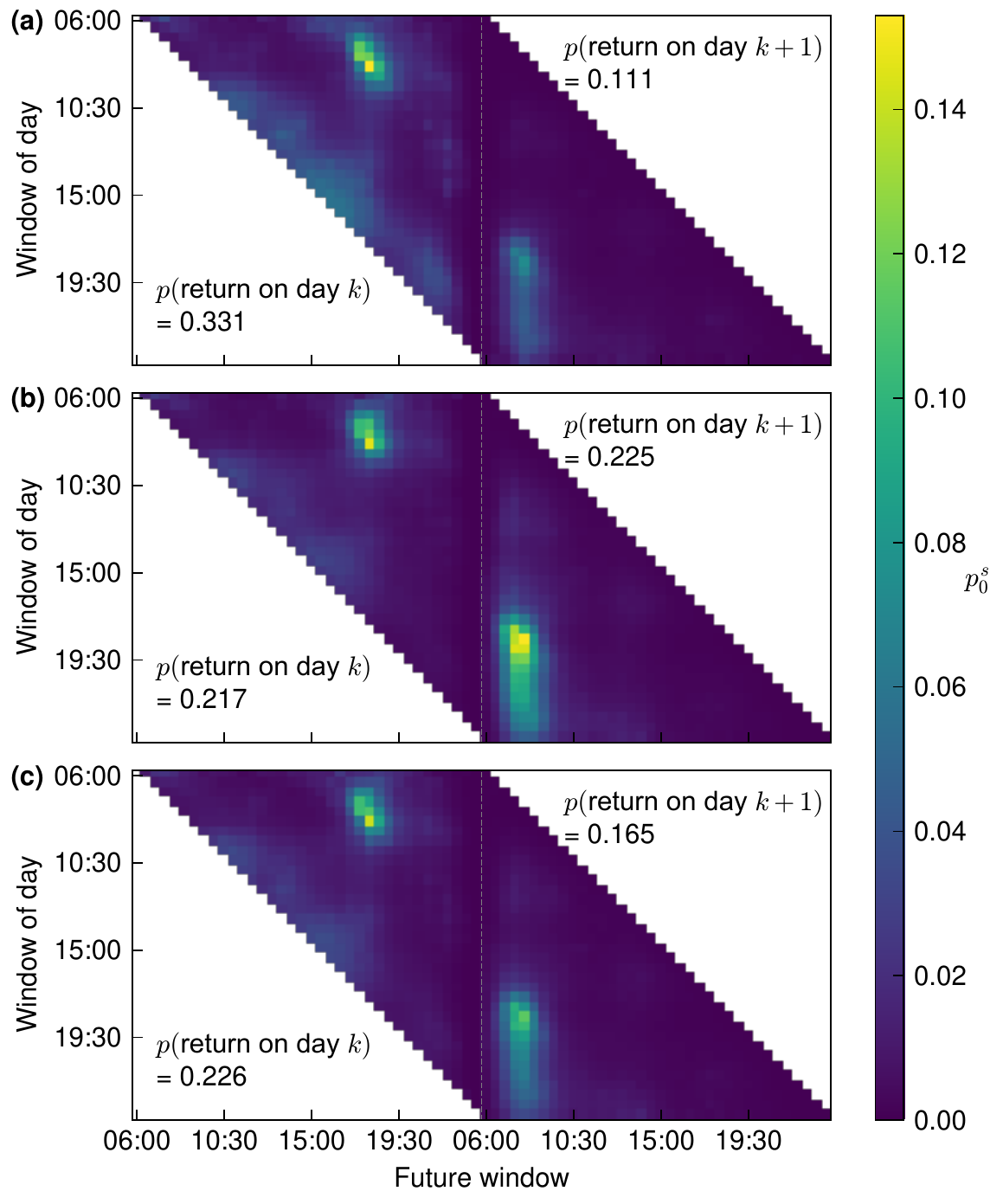}
      \caption{The cluster centroids. (a), (b), and (c) correspond to the cluster centroids of cluster 1, 2, and 3, respectively.}\label{fig:cluster_RPP}
      \end{center}
\end{figure}

The dendrogram for the hierarchical clustering is shown in Figure~\ref{fig:cluster}. We cut the clustering tree at the half-height, which divides the 159 stations into three major clusters (with one station in exception). The cluster centroids (the average RPP for the cluster) for the three clusters are shown in Figure~\ref{fig:cluster_RPP}, where we also show the probabilities of returning on the same day and the next day. From Figure~\ref{fig:cluster_RPP}, we can see cluster 1 corresponds to business-type areas where more passengers return on the same day, such as the Tiyu Xilu station in Figure~\ref{fig:rpp} (a). For cluster 2, the probability of returning on the next day is higher than returning on the same day, exhibiting the feature of residential areas, such as the Luoxi station in Figure~\ref{fig:rpp} (b). Cluster 3 is a combination of cluster 1 and 2, which has a relatively balanced returning flow in both the current and the next day. The bottom panel of Figure~\ref{fig:cluster} shows the $D^s$ values for all stations following the same order as the clustering result. As can be seen, $D^{s}$ values are negative for most stations, confirming the effectiveness of model M2. It should be noted that the effect of $\hat{r}_{t+1}^s$ is different among clusters. The reduction of the SMAPE is the most profound for cluster 1 (with two exceptions, which we will discuss in detail in section~\ref{sec:event}), while visually the least significant for cluster 2. We use the above paired t-test to check if the improvement of M2 compared with M0 is statistically significant. Using the 0.05 significance level, we find the improvements for 23 out of 51 stations (45.1\%) in cluster 1 are significant, 1 out of 41 stations (2.4\%) in cluster 2 are significant, and 9 out of 66 stations (13.6\%)  in cluster 3 are significant. These results suggest the returning flow $\hat{r}_{t+1}^s$ is more effective for the forecast of business-type stations. The reason could be that the duration for work/shop activities is more fixed than the home activity, so the return flow estimation is more accurate for business-type stations.


\subsection{Multi-step ahead forecasting}

Even with strong seasonality, multi-step ahead forecasting is a still challenging task because the errors will accumulate with the rolling forecasting process. A unique advantage of model M2 is that the estimation of returning flow $\hat{r}_{t+L}^s$ suffers less from this error accumulation problem thanks to the long-range dependencies, and even the alighting flow in many steps ago could still dominate the future returning flow. For example, as shown in Figure~\ref{fig:rpp} (a), the alighting flow in the morning (7:00-10:00) plays an important role in determining the returning flow of the evening (17:00-19:00). Therefore, using returning flow as an additional feature in M2  could potentially alleviate the error accumulation problem in multi-step ahead forecasting.

For an $L$-step boarding flow forecasting that predicts $y_{t+L}^s$ by $y_{1:t}^s$, M2 requires a series of returning flow $\hat{r}_{t+1}^s, \ldots, \hat{r}_{t+L}^s$ as input. However, in order to estimate $\hat{r}_{t+L}$, Eq.~\eqref{equ:return_tplus1} requires the alighting flow series $m^s_{t+1:t+L-1}$, which are not available. In this case, we use the average alighting flow at the same window of historical days as the approximation of future alighting flow $m_{t+1}^s \ldots m_{t+L-1}^s$. This approximation for the future alighting flow should only bring minor errors to the estimation of returning flow in Eq.~\eqref{equ:return_tplus1}, since it only contributes to the last $L-1$ components in the summation.

We examine the multi-step ahead forecasting using a time series cross-validation method that is known as ``evaluation on a rolling forecasting origin'' \cite[Chapter~3]{hyndman2018forecasting}. For an $L$-step ahead forecasting, we train a model for each observation in data set D3 using a training set form the first observation in data set D2 to the observation $L$ steps prior to that observation. The error is only evaluated at the $L^{\text{th}}$ step, and the overall error is the average error over the test set.

Table~\ref{tab:multi-step} shows the result of 1, 2, 4, and 6 steps forecasting for M0 and M2.  Compared with M0, M2 offers substantially enhanced forecasting in stations (a), (c), and (d), and the errors increase much slower with the growing step $L$. Especially, in station (a), the RMSE of M0 increases 252.55 (75.6\%) form 1-step forecast to 6-step forecast, the number is only 174.66 (60.4\%) for M2. For the residential (cluster 2) station (b), the effect of M2 in multi-step ahead forecasting is less significant, which validates the different contributions of returning flow among different clusters. Overall, we can see that multi-step ahead forecasting tasks can benefit substantially from the long-range dependencies encoded in M2 and $\hat{r}^s_{t+1}$.

\begin{table}[!ht]
    \centering
    \centering\small
    \caption{The multi-step boarding flow forecasting of four stations by time series cross-validation (30-min resolution).}
      \begin{tabular}{llcccccccc}
  \toprule
    \multirow{2}[4]{*}{Station} & \multirow{2}[4]{*}{Model} & \multicolumn{2}{c}{30 mins (at $L=1$)} & \multicolumn{2}{c}{1 hour (at $L=2$)} & \multicolumn{2}{c}{2 hours (at $L=4$)} & \multicolumn{2}{c}{3 hours (at $L=6$)} \\
\cmidrule{3-10}          &       & RMSE  & SMAPE & RMSE  & SMAPE & RMSE  & SMAPE & RMSE  & SMAPE \\
    \midrule
      \multirow{2}[2]{*}{(a) Tiyu Xilu} & M0    & 334.06 & 12.92\% & 435.16 & 15.30\% & 547.35 & 19.88\% & 586.61 & 19.23\% \\
            & M2    & \textbf{290.43} & \textbf{9.87\%} & \textbf{328.90} & \textbf{12.60\%} & \textbf{403.20} & \textbf{10.82\%} & \textbf{465.09} & \textbf{11.71\%} \\
      \midrule
      \multirow{2}[2]{*}{(b) Luoxi} & M0    & \textbf{75.94} & 10.64\% & 79.53 & 11.70\% & 88.49 & 12.14\% & 90.16 & 12.51\% \\
            & M2    & 76.10 & \textbf{10.60\%} & \textbf{79.43} & \textbf{11.65\%} & \textbf{88.49} & \textbf{12.08\%} & \textbf{89.97} & \textbf{12.38\%} \\
      \midrule
      \multirow{2}[2]{*}{(c) Changshou Lu} & M0    & 97.92 & 10.28\% & 126.65 & 14.42\% & 154.58 & 16.74\% & 168.33 & 15.89\% \\
            & M2    & \textbf{84.80} & \textbf{7.85\%} & \textbf{103.84} & \textbf{8.84\%} & \textbf{129.53} & \textbf{10.37\%} & \textbf{153.12} & \textbf{11.58\%} \\
      \midrule
      \multirow{2}[2]{*}{(d) Huijiang} & M0    & 50.27 & 14.50\% & 54.09 & 15.52\% & 56.16 & 16.22\% & 56.30 & 17.33\% \\
            & M2    & \textbf{39.76} & \textbf{13.95\%} & \textbf{41.13} & \textbf{14.70\%} & \textbf{42.76} & \textbf{15.00\%} & \textbf{43.52} & \textbf{15.69\%} \\
      \bottomrule
      \end{tabular}%
      \label{tab:multi-step}%
\end{table}%

The estimation of the returning trip closely relates to the inter-trip duration. It is thus worth analyzing the effect of time resolution---especially a more refined resolution---to the forecasting performance. We apply a 15-min resolution to further test the impact of the returning flow to the multi-step forecasting. The results are shown in Table \ref{tab:multi-step 15mins}, where the baseline model M0 is ARIMA(2,0,1)(1,1,0)[72]. We can see using the returning flow still greatly alleviates the error accumulation in multi-step forecasting, and the forecasting improvement is the most significant for station (a), (c), and (d), which is consistent with Table \ref{tab:multi-step} and section \ref{sec:short term}.

\begin{table}[!ht]
  \centering\small
  \caption{The multi-step boarding flow forecasting of four stations by time series cross-validation (15-min resolution).}
    \begin{tabular}{llcccccccc}
    \toprule
    \multirow{2}[4]{*}{Station} & \multirow{2}[4]{*}{Model} & \multicolumn{2}{c}{15 mins (at $L=1$)} & \multicolumn{2}{c}{30 mins (at $L=2$)} & \multicolumn{2}{c}{1 hour (at $L=4$)} & \multicolumn{2}{c}{1.5 hour (at $L=6$)} \\
\cmidrule{3-10}          &       & RMSE  & SMAPE & RMSE  & SMAPE & RMSE  & SMAPE & RMSE  & SMAPE \\
    \midrule
    \multirow{2}[2]{*}{(a) Tiyu Xilu} & M0    & 169.57 & 11.63\% & 210.38 & 13.95\% & 248.59 & 16.02\% & 283.43 & 18.86\% \\
          & M2    & \textbf{163.43} & \textbf{10.35\%} & \textbf{190.49} & \textbf{11.44\%} & \textbf{197.12} & \textbf{13.38\%} & \textbf{206.92} & \textbf{13.24\%} \\
    \midrule
    \multicolumn{1}{l}{\multirow{2}[2]{*}{(b) Luoxi}} & M0    & \textbf{41.70} & \textbf{12.45\%} & \textbf{44.31} & 12.79\% & 46.53 & 13.61\% & \textbf{48.68} & 14.29\% \\
          & M2    & 41.72 & 12.46\% & 44.33 & \textbf{12.77\%} & \textbf{46.48} & \textbf{13.58\%} & 48.71 & \textbf{14.25\%} \\
    \midrule
    \multirow{2}[2]{*}{(c) Changshou Lu} & M0    & 52.20 & 12.29\% & 60.30 & 14.26\% & 72.18 & 17.44\% & 79.66 & 19.55\% \\
          & M2    & \textbf{47.79} & \textbf{9.52\%} & \textbf{52.76} & \textbf{10.16\%} & \textbf{60.42} & \textbf{10.45\%} & \textbf{66.06} & \textbf{11.45\%} \\
    \midrule
    \multirow{2}[2]{*}{(d) Huijiang} & M0    & 29.22 & 17.36\% & 30.56 & 17.85\% & 31.90 & 18.14\% & 32.42 & 19.07\% \\
          & M2    & \textbf{25.55} & \textbf{17.21\%} & \textbf{25.90} & \textbf{17.52\%} & \textbf{26.54} & \textbf{17.87\%} & \textbf{26.90} & \textbf{18.63\%} \\
    \bottomrule
    \end{tabular}%
  \label{tab:multi-step 15mins}%
\end{table}%

\subsection{Forecasting under special events}\label{sec:event}

As shown in Figure~\ref{fig:cluster}, M2 is less effective only for a few stations. A main reason is that these stations in general have a large variation in RPP from day to day. For these station, $\hat{r}_{t+1}^s$ will be less accurate and less informative in supporting the forecasting. Therefore, M2 with $\hat{r}_{t+1}^s$ estimated by a universal RPP will not benefit as much, if not more, than M0 and M1.


\begin{figure}[!ht]
      \begin{center}
      \includegraphics[]{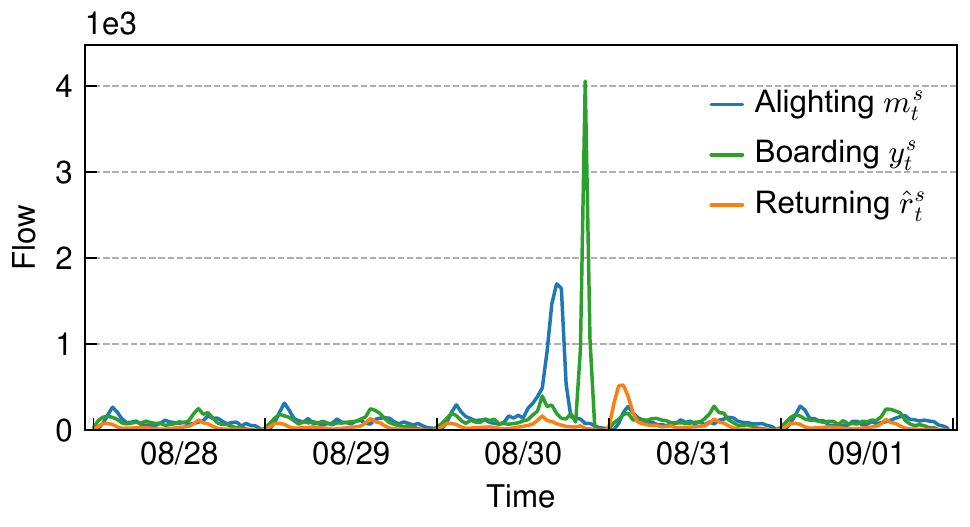}
      \caption{The alighting, boarding, and the estimated returning flow of Luogang station under an event (August 30).}\label{fig:event0}
      \end{center}
\end{figure}

This is particularly the case for forecasting under special events, since the RPP cannot be well estimated using historical data and it will involve large variation in nature. For example, as a metro station next to the Guangzhou International Sports Arena, Luogang often experiences surging demand because of large sports events and concerts. Figure~\ref{fig:event0} shows that the come-and-return dynamics under the special event is very different from normal days. The event on August 30 brought a period of unusually high alighting flow. After the event, the returning passengers caused another peak in the boarding flow. If we adopt a universal RPP estimated by aggregating historical data, we will end up with erroneously distributing the returning flow to the next morning (the morning peak of $\hat{r}_t^s$ on August 31). To address this problem, we propose to build two separate RPPs for normal and event-induced passengers, respectively. For the passengers alight at time $t_a$, the probability of returning at $t$ becomes the weighted sum of the two parts:
\begin{equation}\label{eq:event return probability}
      p^{s}\left(\tau_{\text{boarding}}=t \given \tau_{\text{alighting}}=t_{a} \right) =
      \frac{m^{s,e}_{t_a}}{m^{s}_{t_a}}
      p^{s,e}\left(\tau_{\text{boarding}}=t \given \tau_{\text{alighting}}=t_{a} \right) +
      \frac{m^{s,n}_{t_a}}{m^{s}_{t_a}}
      p^{s,n}\left(\tau_{\text{boarding}}=t \given \tau_{\text{alighting}}=t_{a} \right),
\end{equation}
where we use superscript $e$ and $n$ to denote variables for event and normal conditions, respectively, and thus the alighting flow is $m^{s}_{t_a} = m^{s,e}_{t_a} + m^{s,n}_{t_a}$. When the event-induced alighting flow $m^{s,e}_{t_a}=0$, Eq.~\eqref{eq:event return probability} reduces to the normal RPP. 

In practice, we have a few approaches to estimate the event-induced alighting flow $m^{s,e}_{t_a}$ and RPP under events, such as looking into the passenger flow of the specific gate to the event venue or using the time information of an event. When such information is not available, we propose to apply the following method to estimate the RPP under events (assuming all events in station $s$ follow the same RPP). First, a period with alighting flow larger than a threshold is identified as an event period. For each time window in a day, we use $Q_{3} + 1.5IQR$ as the threshold, where $Q_{3}$ is the third quartile and $IQR$ the interquartile range. Next, the normal RPP can be estimated by the non-event periods. Subtracting the normal alighting (use median) and the normal returning flow from the part identified as event periods, the rest data in event periods are used to estimate the RPP under special events. By separating event and normal flow, we also prevent the normal RPP from being influenced by the event flow.

\begin{figure}[!ht]
      \begin{center}
      \includegraphics[]{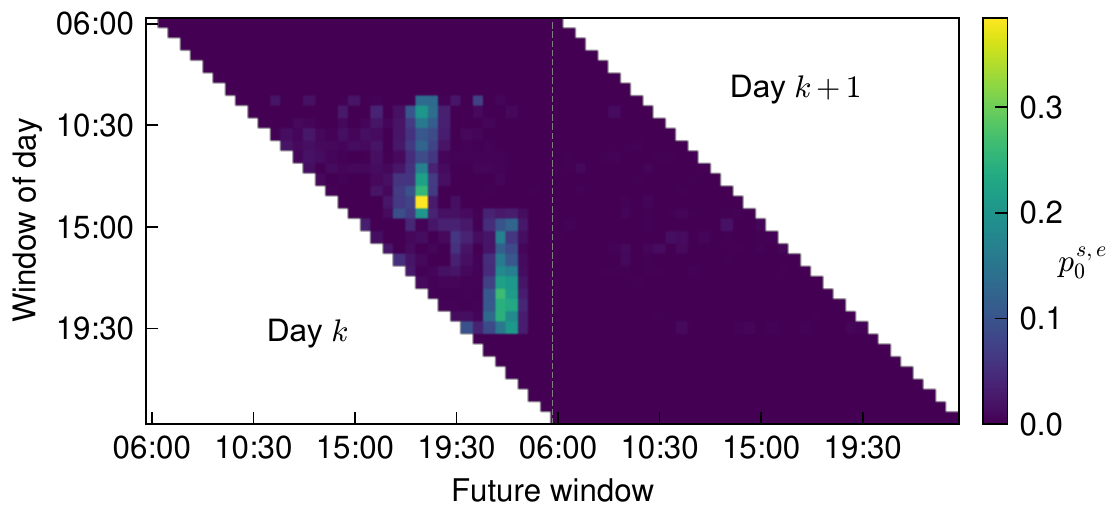}
      \caption{The RPP for passenger flow induced by events in Luogang station. Probabilities are set to zero for time windows without event.}\label{fig:event_RPP}
      \end{center}
\end{figure}

Using the data from July 1 to August 24, 2017 (weekends included), Figure~\ref{fig:event_RPP} shows the event RPP of Luogang station. It is conspicuous that two types of events exist in this venue, one ends in the afternoon and the other ends in the evening. The returning time for each type of event is more concentrated than the alighting time and is mostly on the same day of the alighting time. Although the RPP is estimated from different events (8 days with significant events), the return probability of these events shows regular and predictive patterns.

\begin{figure}[!ht]
      \begin{center}
      \includegraphics[]{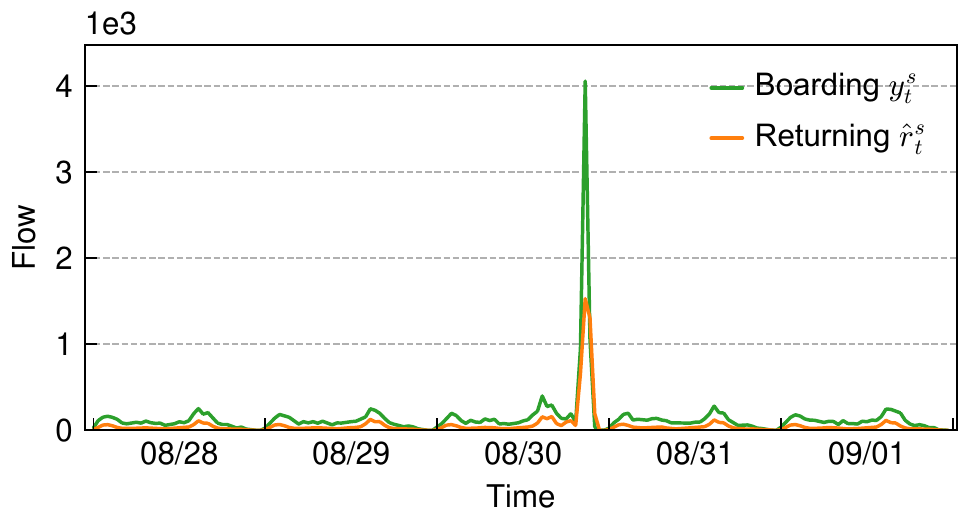}
      \caption{The boarding and the returning flow (using Eq.~\eqref{eq:event return probability}) of Luogang station.}\label{fig:return_flow event}
      \end{center}
\end{figure}

Figure~\ref{fig:return_flow event} shows the re-estimated returning flow of Luogang station. Compared with Figure~\ref{fig:event0}, we can find the returning flow in Figure~\ref{fig:return_flow event} can now correctly reflect the peak and trend of the boarding flow under the special event. Note that the days in Figure~\ref{fig:return_flow event} are different from the days used for event RPP estimation, which shows the come-and-return dynamic of different events in this station follows similar patterns.

Considering that these events are occasional, we use the non-seasonal ARIMA with the order ARIMA$(2,0,1)$ as our baseline model M0. The data separation is the same as section~\ref{sec:data}, except that we use a longer period for the event RPP estimation. We denote by M2' the model that uses the ``adjusted'' returning flow as a covariate. The forecasting results are shown in Table~\ref{tab:result2}. To highlight the forecasting performance under events, we only use event periods (when the boarding flow exceeds $Q_3+1.5IQR$) to calculate the RMSE and the SMAPE, respectively denoted by RMSE$_e$ and SMAPE$_e$. We can see M2' with the ``adjusted'' returning flow has the best performance under all criteria. The results further show that the returning flow offers substantial improvement even for the forecasting under special events. Although the simple ARIMA may not be the best model for a time series with apparent ``outliers'' such as these events, the returning flow could be easily integrated into other models (such as the generalized autoregressive conditional heteroskedasticity, GARCH) for a better prediction.

\begin{table}[!ht]
  \centering \small
  \caption{The boarding flow forecasting under special events for Luogang stations.}
    \begin{tabular}{lrrrrrr}
    \toprule
      \makecell{Model} &  \makecell{RMSE$_e$ \\ (train)} & \makecell{RMSE$_e$ \\ (test)} & \makecell{SMAPE$_e$ \\ (train)} & \makecell{SMAPE$_e$ \\ (test)} & \makecell{Log-likelihood} & AIC \\
        \midrule
        M0    & 392.64 & 1003.29 & 24.55\% & 46.62\%  & -2984.11 & 5978.22 \\
        M1    & 408.15 & 1043.25 & 27.19\% & 45.34\% & -2972.46 & 5956.92 \\
        M2    & 397.14 & 1018.88 & 25.95\% & 46.79\%  & -2909.38 & 5830.76 \\
        M2'   & \textbf{255.46} & \textbf{535.63} & \textbf{23.16\%} & \textbf{36.30\%} & \textbf{-2821.29} & \textbf{5654.58} \\
        \bottomrule
    \end{tabular}%
    \label{tab:result2}%
\end{table}%

\section{Conclusions and discussion}\label{sec:conclusion}

In this paper, we propose a new framework for forecasting passenger flow time series in metro systems. In contrast to some previous studies that capture temporal dynamics in a data-driven way, we try to incorporate the generative mechanisms rooted in travel behavior into modeling passenger flow time series. For that purpose, we introduce returning flow as a new covariate/feature into standard time series models. This returning flow is estimated as the expected returning boarding demand given previous alighting trips; thus, it encodes the causal structure and long-range dependencies in passenger flow data. We estimate the return probability by aggregating historical data, thereby working around the sensitivity issues and privacy concerns accompanying individual-based data and models as in \citet{zhao2018individual}. We examine the proposed framework on a real-world passenger flow data set collected from Guangzhou metro in China. The proposed framework with the returning flow demonstrates superior performance in all three tested scenarios, namely one-step ahead forecasting, multi-step ahead forecasting, and forecasting in special events. And we found the returning flow is more useful for the boarding demand forecast of business-type stations, where most returning trips are within the same day. On the contrary, the model does not bring much improvement for residential stations. This result suggests that ``home'' activity duration demonstrates a higher variance than that of ``work'' activities. In addition, the experiments in \ref{appendix:A} show the returning flow also improves the forecast of machine learning models like SVR and MLP. In fact, the returning flow (as a covariate) and the idea of regularity-based long-range dependency can be used in a diverse range of prediction models (e.g., time series model, machine learning models, Deep learning).

There are several directions for future research. First, this study assumes that both the boarding and the alighting time series are available (i.e., both tapping-in and tapping-out are registered by the smart card system), while metro in some cities may have a tapping-in only system. In this case, one should integrate a destination inference model \citep[see e.g.,][]{barry2002origin, trepanier2007individual, cheng2020probabilistic} into the proposed framework. Estimating the returning probability by other data sources, such as survey, Bluetooth, and call detail records, is also worth exploring. Second, the come-and-return pattern and RPP of a station may change over time. How to detect the pattern changes \citep{zhao2018detecting} and develop time-varying models should be studied. The current constant RPP is a simplified assumption; further utilizing the returning flow's auto-correlation is a possible approach to improve the returning flow estimation. There is still space to improve the model by advanced statistical time series and deep learning-based sequence models. Third, the models developed in the current framework are station-specific, while travel behavior regularity is ubiquitous and different stations may share similar patterns. Therefore, the RPP formulation can be generalized using parametric model, approximation, and dimensionality reduction techniques such as principal component analysis and matrix/tensor factorization \citep{sun2016understanding} to extract common patterns for RPPs across different stations. This can be particularly useful when only limited data are available. Fourth, we should relax the assumption that the returning flow must have the same boarding station as the previous alighting station, because there could be multiple destinations or multiple transportation modes in one’s activity chain \citep{bowman2001activity}. It is possible to improve and extend the forecasting by multi-modal trips since the metro system is often combined with other transportation modes; a route/mode choice model can be integrated into the multi-modal forecasting framework. Sufficient data is a prerequisite for this direction. Lastly, we can extend this framework to other transport modes with non-random and chained travel patterns, such as private vehicles, taxis, and ride-hailing services. This paper also sheds new light on other behavior-driven demand forecasting problems, in which the causal structure and the long-range dependencies play a substantial role. For instance, integrating purchasing behavior into the demand forecasting of retail products.

\section*{Acknowledgement}
This research is supported by the Natural Sciences and Engineering Research Council (NSERC) of Canada, Mitacs, exo.quebec (https://exo.quebec/en), NSFC-FRQSC Research Program on Smart Cities and Big Data, the Institute for Data Valorisation (IVADO), and the Canada Foundation for Innovation (CFI).

\bibliographystyle{elsarticle-harv}
\bibliography{references} 

\appendix

\section{Experiments in other models}\label{appendix:A}
To further test if the returning flow can also improve other ridership forecast models, we use two popular machine learning models--Support Vector Regression (SVR) and Multi-Layer Perceptron (MLP)--to repeat the experiment in section~\ref{sec:short term}. We rescale data to $[0,1]$ by min-max normalization as a preprocessing, and we use the scikit-learn python package to implement these models.

The SVR model is similar to the previous work \citep{tang2018forecasting} except we have no external features like the weather. The input features for M0 are the boarding flow at time $t$ and $t-1$, and 36 dummy variables representing the time of a day. We add $r^{s}_{t}$ to M1 and $\hat{r}_{t+1}^s$ to M2 as an additional feature. We tune hyper-parameters by cross-validation using M0 and select $C = 0.274$ and $\varepsilon=0.016$; other hyper-parameters are the default setting of the scikit-learn package. The forecast results of the four stations by SVR are shown in Table~\ref{tab:svr-forecast} and the significance tests are shown in Table~\ref{tab:svr-ttest}.

\begin{table}[htbp]
      \centering\small
      \caption{The one-step boarding flow forecasting of four stations by SVR}
        \begin{tabular}{llrrrr}
        \toprule
        Stations & Model &  \makecell{RMSE \\ (train)} & \makecell{RMSE \\ (test)} & \makecell{SMAPE \\ (train)} & \makecell{SMAPE \\ (test)} \\
        \midrule
        \multirow{3}[2]{*}{(a) Tiyu Xilu} & M0    & 288.00 & 337.02 & 11.27\% & 11.60\% \\
              & M1    & 286.54 & 337.75 & 11.59\% & 11.76\% \\
              & M2    & \textbf{280.32} & \textbf{306.79} & \textbf{11.12\%} & \textbf{11.45\%} \\
        \midrule
        \multirow{3}[2]{*}{(b) Luoxi} & M0    & \textbf{64.61} & 80.66 & \textbf{11.30\%} & 13.41\% \\
              & M1    & 65.81 & 78.45 & 11.37\% & \textbf{13.39\%} \\
              & M2    & 64.86 & \textbf{78.34} & 11.47\% & 13.53\% \\
        \midrule
        \multirow{3}[2]{*}{(c) Changshou Lu} & M0    & 72.80 & 111.81 & \textbf{6.00\%} & 9.72\% \\
              & M1    & 72.91 & 109.99 & 6.13\% & 9.51\% \\
              & M2    & \textbf{65.71} & \textbf{103.83} & 6.06\% & \textbf{9.33\%} \\
        \midrule
        \multirow{3}[2]{*}{(d) Huijiang} & M0    & 31.95 & 45.37 & 13.53\% & 15.75\% \\
              & M1    & 31.98 & 44.46 & 13.61\% & 15.83\% \\
              & M2    & \textbf{31.64} & \textbf{42.53} & \textbf{13.46\%} & \textbf{15.75\%} \\
        \bottomrule
        \end{tabular}%
      \label{tab:svr-forecast}%
    \end{table}%

\begin{table}[htbp]
\centering\small
\caption{Paired t-test p-values for SVR absolute forecast errors.}
\begin{threeparttable}
\begin{tabular}{lrrrr}
      \toprule
            & (a) Tiyu Xilu & (b) Luoxi & (c) Changshou Lu & (d) Huijiang \\
      \midrule
      \makecell{$\mathrm{H}_0: \mu\left(|\varepsilon_{\mathrm{M2}}| -|\varepsilon_{\mathrm{M0}}|\right) = 0$\\ $\mathrm{H}_a: \mu\left(|\varepsilon_{\mathrm{M2}}| -|\varepsilon_{\mathrm{M0}}|\right) < 0$} & $<0.001^{*}$ & 0.403 & $<0.001^{*}$ & $0.022^{*}$ \\
      \midrule
      \makecell{$\mathrm{H}_0: \mu\left(|\varepsilon_{\mathrm{M2}}| -|\varepsilon_{\mathrm{M1}}|\right) = 0$\\ $\mathrm{H}_a: \mu\left(|\varepsilon_{\mathrm{M2}}| -|\varepsilon_{\mathrm{M1}}|\right) < 0$} & $<0.001^{*}$ & 0.768 & $<0.001^{*}$ & $0.011^{*}$ \\
      \bottomrule
      \end{tabular}%
\begin{tablenotes}
      \item $^{*}$ Significant at 0.05 level.
\end{tablenotes}
\end{threeparttable}
\label{tab:svr-ttest}%
\end{table}%

The MLP uses the same features as the SVR. We tune hyper-parameters by cross-validation using M0 and select the hidden layer size to be 150 and use the identity activation function; other hyper-parameters are the default setting of the scikit-learn package. The forecast results of the four stations by MPL are shown in Table~\ref{tab:mlp-forecast} and the significance tests are shown in Table~\ref{tab:mlp-ttest}.

\begin{table}[htbp]
      \centering\small
      \caption{The one-step boarding flow forecasting of four stations by MLP}
        \begin{tabular}{llrrrr}
        \toprule
        Stations & Model &  \makecell{RMSE \\ (train)} & \makecell{RMSE \\ (test)} & \makecell{SMAPE \\ (train)} & \makecell{SMAPE \\ (test)} \\
        \midrule
        \multirow{3}[2]{*}{(a) Tiyu Xilu} & M0    & 312.23 & 332.87 & 12.08\% & 13.52\% \\
              & M1    & 309.10 & 333.81 & 11.12\% & 13.08\% \\
              & M2    & \textbf{296.60} & \textbf{307.57} & \textbf{10.68\%} & \textbf{12.45\%} \\
        \midrule
        \multirow{3}[2]{*}{(b) Luoxi} & M0    & \textbf{59.86} & \textbf{74.89} & \textbf{9.92\%} & \textbf{11.68\%} \\
              & M1    & 71.42 & 82.27 & 13.57\% & 15.56\% \\
              & M2    & 72.66 & 80.54 & 14.30\% & 15.92\% \\
        \midrule
        \multirow{3}[2]{*}{(c) Changshou Lu} & M0    & 74.03 & 108.54 & \textbf{7.28\%} & 13.37\% \\
              & M1    & 77.38 & 105.13 & 9.19\% & 13.60\% \\
              & M2    & \textbf{65.50} & \textbf{95.31} & 7.41\% & \textbf{12.35\%} \\
        \midrule
        \multirow{3}[2]{*}{(d) Huijiang} & M0    & 32.50 & 43.08 & \textbf{12.62\%} & 15.59\% \\
              & M1    & 34.64 & 43.85 & 14.03\% & 15.59\% \\
              & M2    & \textbf{31.39} & \textbf{35.86} & 14.02\% & \textbf{15.20\%} \\
        \bottomrule
        \end{tabular}%
      \label{tab:mlp-forecast}%
    \end{table}%

\begin{table}[htbp]
      \centering\small
      \caption{Paired t-test p-values for MLP absolute forecast errors.}
      \begin{threeparttable}
        \begin{tabular}{lrrrr}
        \toprule
              & (a) Tiyu Xilu & (b) Luoxi & (c) Changshou Lu & (d) Huijiang \\
        \midrule
        \makecell{$\mathrm{H}_0: \mu\left(|\varepsilon_{\mathrm{M2}}| -|\varepsilon_{\mathrm{M0}}|\right) = 0$\\ $\mathrm{H}_a: \mu\left(|\varepsilon_{\mathrm{M2}}| -|\varepsilon_{\mathrm{M0}}|\right) < 0$}  & $<0.001^{*}$ & $0.999$ & $<0.001^{*}$ & $0.031^*$ \\
        \midrule
        \makecell{$\mathrm{H}_0: \mu\left(|\varepsilon_{\mathrm{M2}}| -|\varepsilon_{\mathrm{M1}}|\right) = 0$\\ $\mathrm{H}_a: \mu\left(|\varepsilon_{\mathrm{M2}}| -|\varepsilon_{\mathrm{M1}}|\right) < 0$} & $<0.001^{*}$ & 0.607 & $0.003^*$ & $<0.001^{*}$ \\
        \bottomrule
        \end{tabular}%
      \begin{tablenotes}
            \item $^*$ Significant at 0.05 level.
      \end{tablenotes}
\end{threeparttable}
      \label{tab:mlp-ttest}%
\end{table}%

In Table~\ref{tab:svr-forecast} and Table~\ref{tab:mlp-forecast}, M2 has lower forecast RMSE and SMAPE in the test set than M0 and M1 for stations (a)(c)(d). The hypothesis tests in Table~\ref{tab:svr-ttest} and Table~\ref{tab:mlp-ttest} for these stations also show that the absolute forecast error of M2 is less than M0 and M1. On the other hand, the effect of using the returning flow is not significant for station (b)---a residential type station. In summary, the returning flow can improve the boarding flow forecast of SVR and MLP, and the improvement is more significant for business-type stations with more passengers return on the same day (i.e., station (a)(c)(d)). These results are consistent with the SARIMA model.

Finally, we apply SVR and MLP to all stations and use the paired t-test described in section~\ref{sec:short term} to test if the improvement of M2 is significant compared with M0. Results are shown in Table.~\ref{tab:ttest-all}, where the clusters are the same as Fig.~\ref{fig:cluster}. The results of different models are consistent, the proposed returning flow is more effective for business-type stations.

\begin{table}[htbp]
      \centering\small
      \caption{The number of significant stations in the paired t-test between M2 and M0 (0.05 significance level).}
      \begin{tabular}{cccc}
      \toprule
         & \makecell{Cluster 1 \\ (business-type)} & \makecell{Cluster 2\\(residential-type)}  & \makecell{Cluster 3\\(combined-type)} \\
      \midrule
        Number of stations & 51 & 41 & 66 \\
        SARIMA & 23 (45.1\%) & 1 (2.4\%) & 9 (13.6\%) \\
        SVR   & 22 (43.1\%) & 4 (9.8\%) & 13 (19.7\%) \\
        MLP   & 29 (56.8\%) & 4 (9.8\%) & 13 (19.7\%) \\
        \bottomrule
        \end{tabular}%
      \label{tab:ttest-all}%
\end{table}%

\end{document}